\documentclass[14pt]{article}
\usepackage{arxiv}
\usepackage[utf8]{inputenc} % allow utf-8 input
\usepackage[T1]{fontenc}    % use 8-bit T1 fonts
\usepackage{hyperref}       % hyperlinks
\usepackage{url}            % simple URL typesetting
\usepackage{booktabs}       % professional-quality tables
\usepackage{amsfonts}       % blackboard math symbols
\usepackage{nicefrac}       % compact symbols for 1/2, etc.
\usepackage{amsfonts}
\usepackage{graphicx}
\usepackage[table,xcdraw]{xcolor}
\usepackage{gensymb}
\newtheorem{theorem}{Theorem}
\newtheorem{definition}{Definition}
\title{Towards Emotion Recognition: A Persistent Entropy Application}

\author{
  Rocio Gonzalez-Diaz\ \\
  Dept. of Applied Mathematics I \\
  University of Seville\\
  \texttt{rogodi@us.es} \\
   \And
 Eduardo Paluzo-Hidalgo \\
 Dept. of Applied Mathematics I \\
  University of Seville\\
  \texttt{epaluzo@us.es} \\
  \And
  Jos\'e F. Quesada \\
 Dept. of Computer Science and Artificial Intelligence \\
  University of Seville\\
  \texttt{jquesada@us.es} \\
}

\begin{document}
\maketitle

\begin{abstract}
Emotion recognition and classification is a very active area of research. In this paper, we present a first approach to emotion classification 
using persistent entropy and support vector machines.
%topological
%by topology 
%tools. 
%More concretely, a topology-based model is applied 
%in order 
A topology-based model is applied to obtain %$\mathbb{R}$
a single real number from each raw signal.
%values from raw signals 
%in order to differentiate
%to do emotion classification 
%between $8$ 
%different 
%emotions (calm, happy, sad, angry, fearful, disgust and surprised). 
%Then, 
These data are used as input of a support vector machine to classify signals into $8$ different emotions (calm, happy, sad, angry, fearful, disgust and surprised).
\end{abstract}

% keywords can be removed

\section{Introduction}

Emotion recognition is not a trivial task and different approaches %can be seen in literature 
have been explored so far (see for example \cite{soundresearch}).  
Additionally, its applications are really important, such as
gathering and processing satisfaction feedback in customers' services, generating statistical studies over a population, using  emotion recognition to improve spoken language understanding during a conversation. Furthermore, it can help in human interaction as in KRISTINA project\footnote{\url{http://kristina-project.eu/en/}}, where emotion recognition is applied in order to help the interaction between health professionals and migrated patients. Among the different theories about emotions proposed in the specialized literature, we follow the model described in \cite{ortony1990s} and  \cite{russell1980circumplex}, where a discrete theory of emotions is given, %differing 
differentiating
several basic groups of emotions (neutral, happy, sad and surprised) and organizing them in a spatial model. In \cite{Ververidis06emotionalspeech} a review of different emotional speech recognition techniques can be consulted. 

 Topological data analysis is a well substantiated field useful to extract information from data (see \cite{doi:10.1146/annurev-statistics-031017-100045}). Concretely, a recent tool in this area called {\em persistent entropy}  has been successfully applied to distinguish discrete piecewise-linear functions (see \cite{RUCCO2017130}). 
 
In this paper, persistent entropy %application is intended, trying 
is used to model arousal (i.e., 
emotional state)
%the state of emotion) 
and emotion recognition as follows. 
First, speech signals are considered as piecewise linear functions. Second, persistent entropy is computed from the lower-star filtration obtained from these functions.
%to recognize emotions in speech. 
This persistent entropy embedding can be considered as a summary of the features that appear in raw signals, as intensity and intonation. 
%Furthermore, 
The stability theorem for persistent entropy
computed from lower-star filtrations
\cite{RUCCO2017130} guarantees right comparison between signals and robustness against noise.
Finally, a support vector machine is used to classify emotions via
%their 
persistent entropy values. As far as our knowledge, no topology approaches have been previously applied to emotion recognition. 

This paper is organized as follows: Basic emotion theory,
%%%%%%%%%%%%%%%%%%%%%%%%%%%%%%%%%%%%%%%%%%%%5
%ROCIO: lo he quitado porque suena repetitivo
% that base our experiments is provided, 
%%%%%%%%%%%%%%%%%%%%%%%%%%%%%%%%%%%%%%%%%%%%%%%
the notions of persistent homology and persistent entropy, and machine learning knowledge required for the model are introduced in Section \ref{sec:background}. In Section \ref{sec:methods}, the methodology followed in the experiments is explained. 
Results obtained  %giving 
from different training approaches are %commented and 
shown in Section \ref{sec:experiments}. 
Finally, %Then, 
Section \ref{sec:conclusions} provides conclusions and future work ideas.

\section{Background}\label{sec:background}

In this paper, different tools are mixed up in order to propose a
unified and coherent framework for emotion classification. In this section, the basic concepts about acoustics, topology, machine learning and statistics are introduced.

\medskip

\noindent {\bf Acoustic and Psychoacoustic Features.} Emotions
%Feelings 
constitute the main field largely studied by psychologists. Following
%Basing on 
\cite{russell1980circumplex}, we consider that emotions can be modeled spatially in a circle, being arousal and valence their main characteristic features. 
Accordingly, prosodic attributes of speech \cite{Globerson2013} are strongly related with emotion recognition. 
This research area takes into account several features of speech, in conjunction with gesticulation of the speaker. Some of those features are: pitch signal, number of harmonics, vocal tract, and speech energy. 

Along this paper, just the physical features of the acoustic signal along with the processing results available from this signals  (such as the contour of speech signal which is a feature affected by the arousal of the speaker),
will be taken into account. 
%\ro{We will use pitch, energy and duration as main information features.} 
The inclusion of visual features will be proposed 
in Section \ref{sec:conclusions}
as a natural continuation of this research. Sentences will be processed, assuming that certain attributes, like the fundamental frequency, intensity and duration, of a sound are meaningful for emotion production and recognition. These attributes are encapsulated under the notion of prosody. Depending on the prosodic pattern, a sentence can have very different emotional features. For example, happiness is linked usually with large fundamental frequency and, loudness, in contrast with sadness, normally related to the opposite. For further explanations about psychoacoustics, \cite{Howard:2000:AP:556750} can be consulted.

%The unit we will manipulate and that is subject of study are sentences of natural speech. These sentences are sound waves. A sound wave are physically composed by compression and decompression of the molecules of the air.
%%%%%%%%%%%%%%%%%%%%%%%%%%%%%%%%%%%%%%%%5
%ROCIO: Esta definicion no se usa
%\begin{definition}
%[Utterance] An utterance is a speech segment corresponding to %a word or a phrase.
%\end{definition}
%%%%%%%%%%%%%%%%%%%%%%%%%%%%%%%%%%%%%%

%%%%%%%%%%%%%%%%%%%%%%
%QUé son formants?
%%%%%%%%%%%%%%%%%%%%%5

%For instance, 

%%%%%%%%%%%%%%%%%%%%%%%%%55
%ROCIO: esto no lo entiendo:
%Here we expose these concepts. 
%%%%%%%%%%%%%%%%%%%%
%However, intensity of the signal is our main subject of research in this first approach.

%%%%%%%%%%%%%%%55
%ROCIO:speech signal es lo mismo que intensity of signal? 
%no creo que haga falta ponerlo como definicion sino ponerlo sobre la marcha y siempre llamarlo igual
%%%%%%%%%%%%%%%%%%%%%%5

In the literature, some emotion classification techniques have been proposed (see \cite{YANG20101415}). Some of them employ prosody contours information of speech in order to recognize emotions, 
as, for example:
artificial neural networks, the multichannel hidden Markov model, and the mixture of hidden Markov models. 
For a further approximation to paralinguistic theory see \cite{schuller2013computational}.

\medskip

\noindent {\bf Topology background.} 
%model, 
%Persistent entropy is the main tool that will be used
Topological data analysis (TDA) studies the {\it shape of data}.
In our case, we 
%are trying to 
apply topological data analysis tools to distinguish between piecewise linear function shapes. For an introduction to topological data analysis, \cite{Edelsbrunner10} can be consulted.

Persistent entropy is the main tool from TDA that will be used in this paper. It sums up   persistent homology information which  ``measures" homological features of shapes and of functions. 

Informally, homology provides
%studies 
the number of $n$-dimensional holes, called the $n$-th Betti numbers and denoted by $\beta_n$. 
Intuitively,
$\beta_0$ is
%measures 
the number of connected components, $\beta_1$ the number of tunnels and $\beta_2$ the number of cavities.
However, for dimensions higher than $2$, we lose the intuition about what a hole is.

\begin{definition}[Betti number, informal, \cite{Bredon}]
If $X$ is a topological space, then $H_n(X)\simeq \mathbb{Z}^{\beta_n}$ is called the $n$-th homology group of $X$ if the power $\beta_n$ is the number of 
independent $n$-dimensional
'holes' %of dimension $n$ 
in $X$. 
%Note that $\beta_0$ is the number of separate connected components. 
We call $\beta_n$ the $n$-th Betti number of $X$. Finally, the homology of $X$ is defined as $H(X)= \{H_n(X)\}^\infty_{n=0}$.
\end{definition}
%
%ro{In practice, the way to compute homology is as follows: Suppose we have an object structured as a simplicial complex. A collection $K$ of simplices (vertices, edges, triangles and their higher dimensional counterparts) is a simplicial complex if it satisfied that:
%satisfying that:
%\begin{itemize}
%    \item[(1)] If a simplex is in $K$, then the simplices in its boundary are in $K$, where, for example, the boundary of an edge is the set of its two endpoints (vertices) and the boundary of a triangle is the set of its three sides (edges).
%    \item[(2)] If two simplices intersect, they intersect at their boundaries.
%\end{itemize}
% }
%
%\ro{Definir: boudary map. Homology groups as quotient groups. }

Observe that the concept of homology is not useful in practice. For example, suppose a dataset $V$ of $10$ points sampling a circumference. We expect that $H_0(V)\simeq \mathbb{Z}$ since a circumference has one connected component. However, the exact $0$-th homology of $V$ is $\mathbb{Z}^{10}$.
%
%When we are dealing with some data, and we want to consider its homology, we are referring to its underling homology. By this, we mean that, if, for example, we want to compute $H_0$ of 10 points sampling a circumference, we expect that $H_0$ is equal to one connected component. However, the exact homology for $0$ dimensions is equal to $10$ in this example. 
%
Therefore, we need a tool to compute the homology of the underlying space sampled by a dataset. 
Following this idea, Edelsbrunner et al. \cite{Edelsbrunner10} introduced the concept of persistent homology together with an efficient algorithm and its visualization as a persistence diagram. Carlsson et al. \cite{zomorodian2005computing} reformulated and extended the initial definition and gave an equivalent visualization method called persistence barcodes.

Given a dataset $V$ and a simplicial complex $K$ constructed from it, 
persistent homology measures homology by a filtration during time, obtaining births and deaths of each homology class ('hole'). Consequently, those classes that persist are better candidates to be representatives of the homology of the underlying space.

\begin{definition}[Abstract simplicial complex]
        Let $V$ be a finite set. A family $K$ of subsets of $V$ is an abstract {\emph simplicial complex} if for every subsets $\sigma \in K$ and $\mu \subset V$, we have that $\mu \subset \sigma$ implies $\mu \in K$.
        A subset in $K$ of $m+1$ elements of $V$ is called a $m$-simplex and $V$ is called the set of vertices of $K$.
    \end{definition}

\begin{definition}[Filtration]
Given a set $V$ and a simplicial complex $K$ constructed from it, a filtration is a finite increasing sequence of simplicial complexes:
$$\emptyset= K_0 \subset K_1\subset K_2\subset \dots \subset K_n=K $$
\end{definition}

\noindent A particular filtration that will be used in this paper is the lower-star filtration.

\begin{definition}[Lower-star filtration \cite{Edelsbrunner10}]
 Let $K$ be a simplicial complex with real (distinct) values specified on the set $V$ of all the vertices in %$S$ 
 $K$.
 %$f$ is assumed generic,i.e, vertices have distinct function values. 
 %Then, the 
 Since vertices have distinct function values, then they
can be ordered %by increasing 
incrementally:
%function value as
$$f(u_1)<f(u_2)<\dots < f(u_n).$$
%Finally, 
The lower star of $u_i$ %can be computed which 
is the subset of simplices of $K$ for which $u_i$ is the vertex with maximum function value,
$$K_i = \{\sigma \in K \ : \ \mbox{for all vertex} \ v \mbox{ of }  \sigma \Rightarrow f(v) \le f(u_i) \}. $$
\end{definition}
\label{def:starfiltr}

\begin{center}
\begin{table}[]
    \centering
    \begin{tabular}{|c c|}
    \hline
       \includegraphics[width = 0.45 \linewidth]{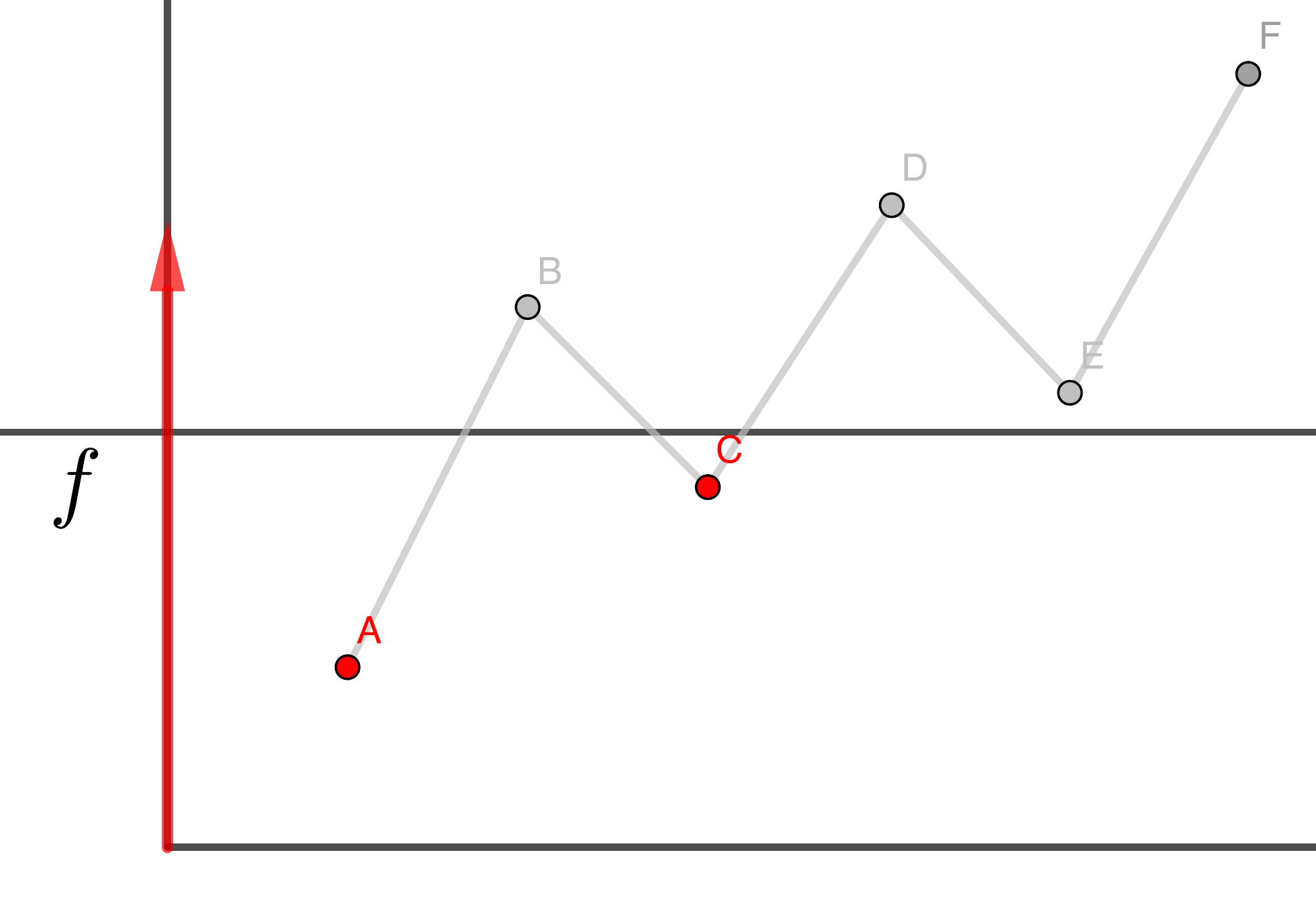}\hspace{10mm} &\includegraphics[width = 0.3 \linewidth]{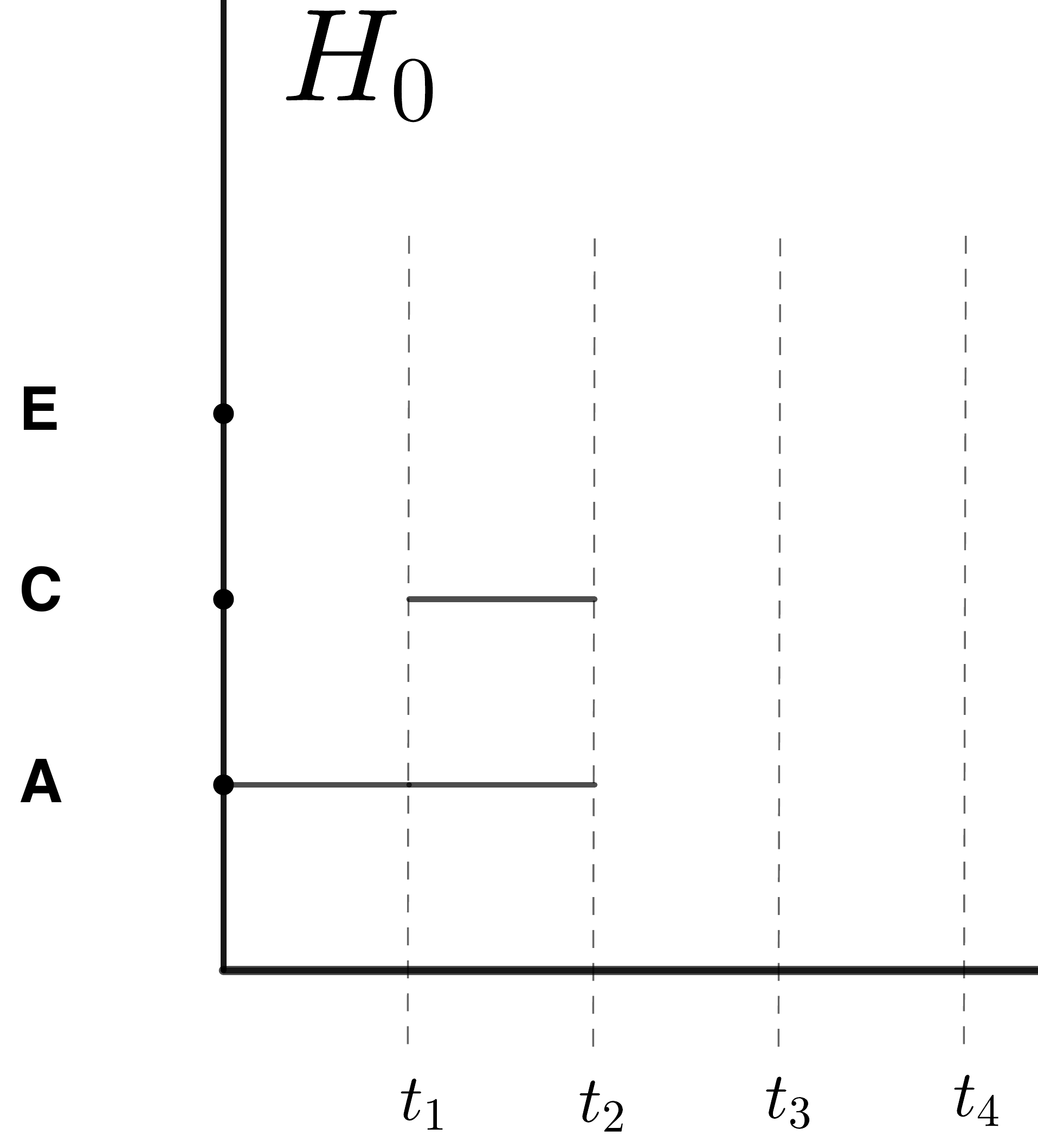} \hspace{5mm} \\\hline 
        \includegraphics[width = 0.45 \linewidth]{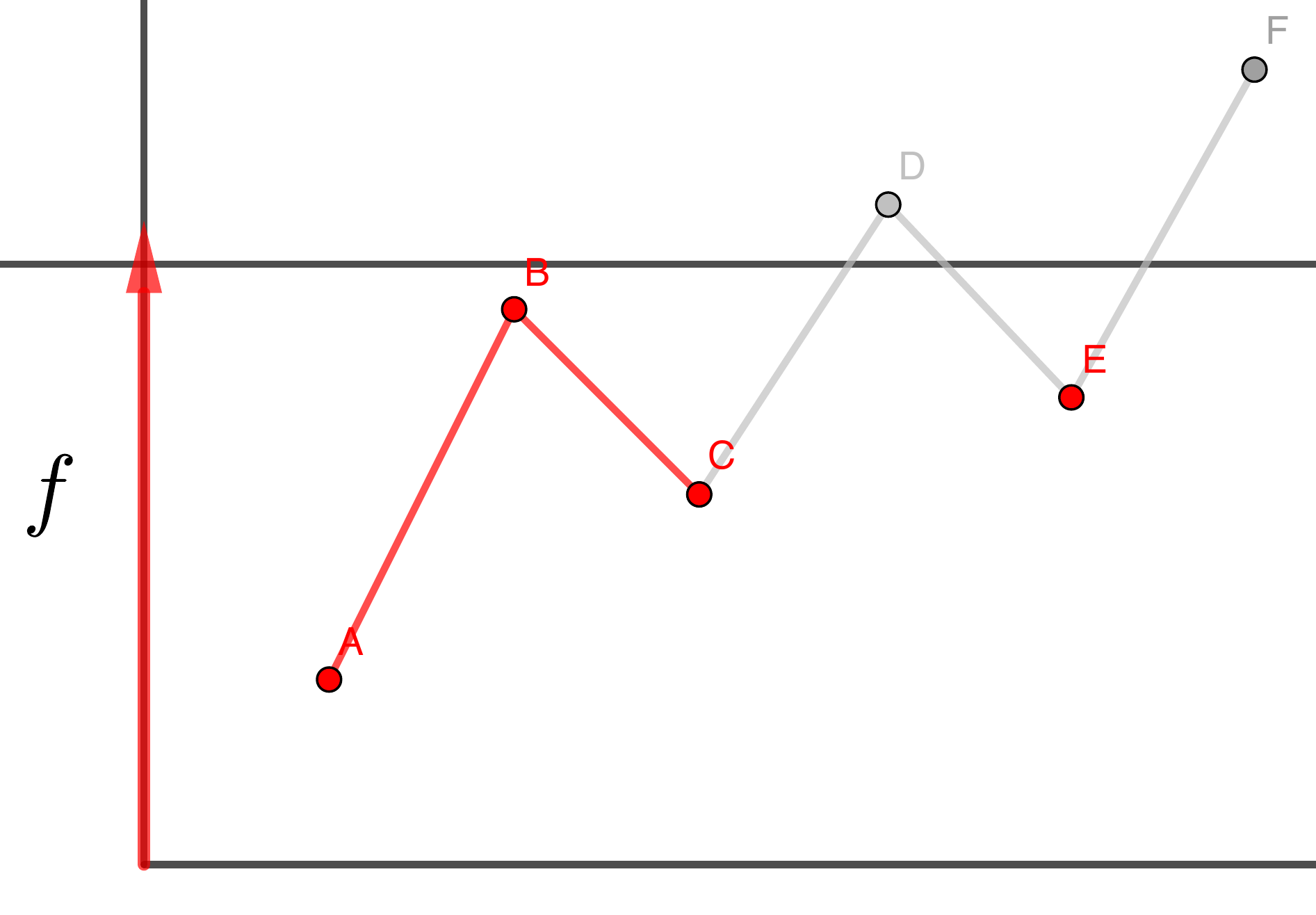} \hspace{10mm} &  \includegraphics[width = 0.3 \linewidth]{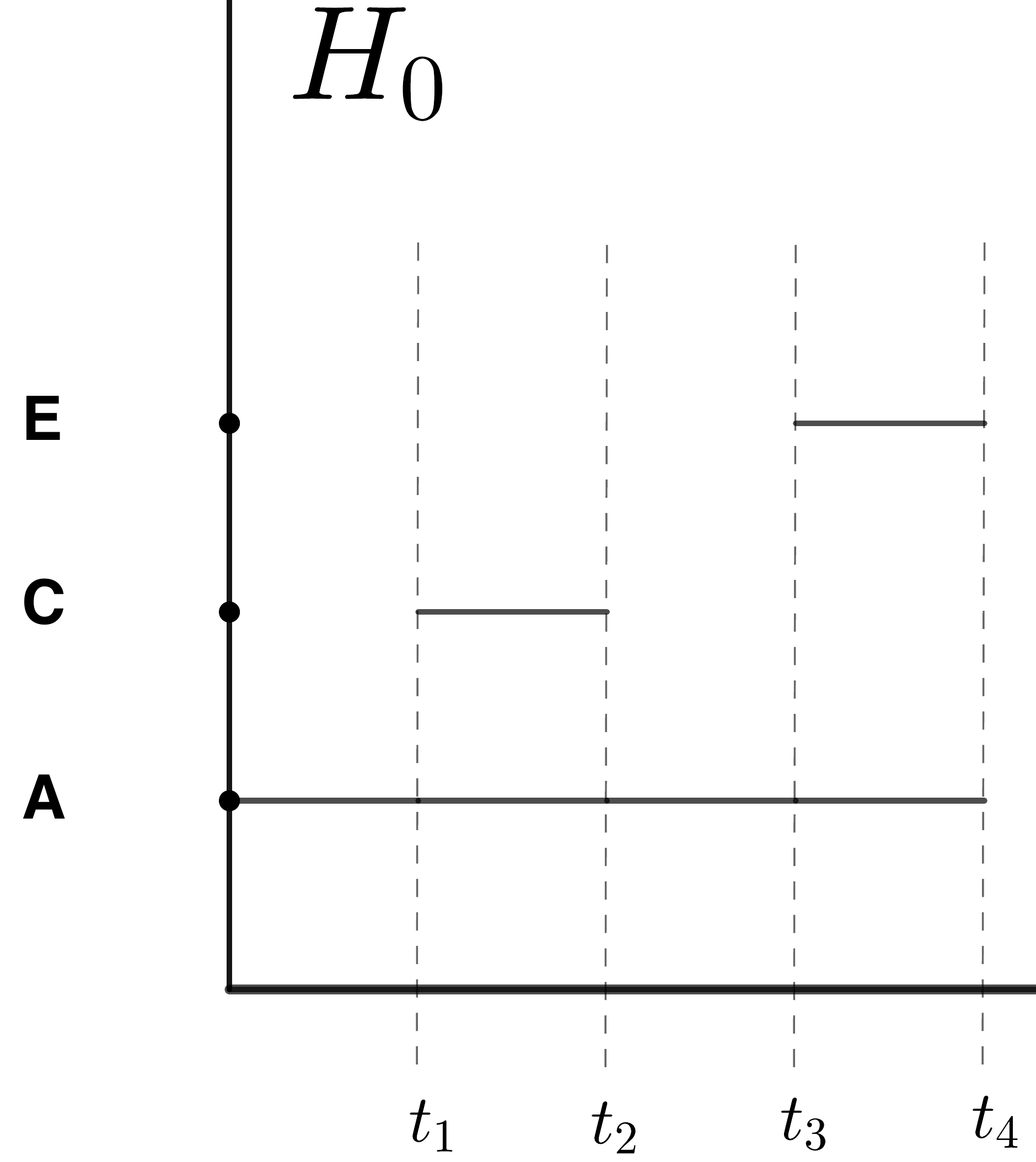}  \hspace{5mm} \\\hline 
        \includegraphics[width = 0.45\linewidth]{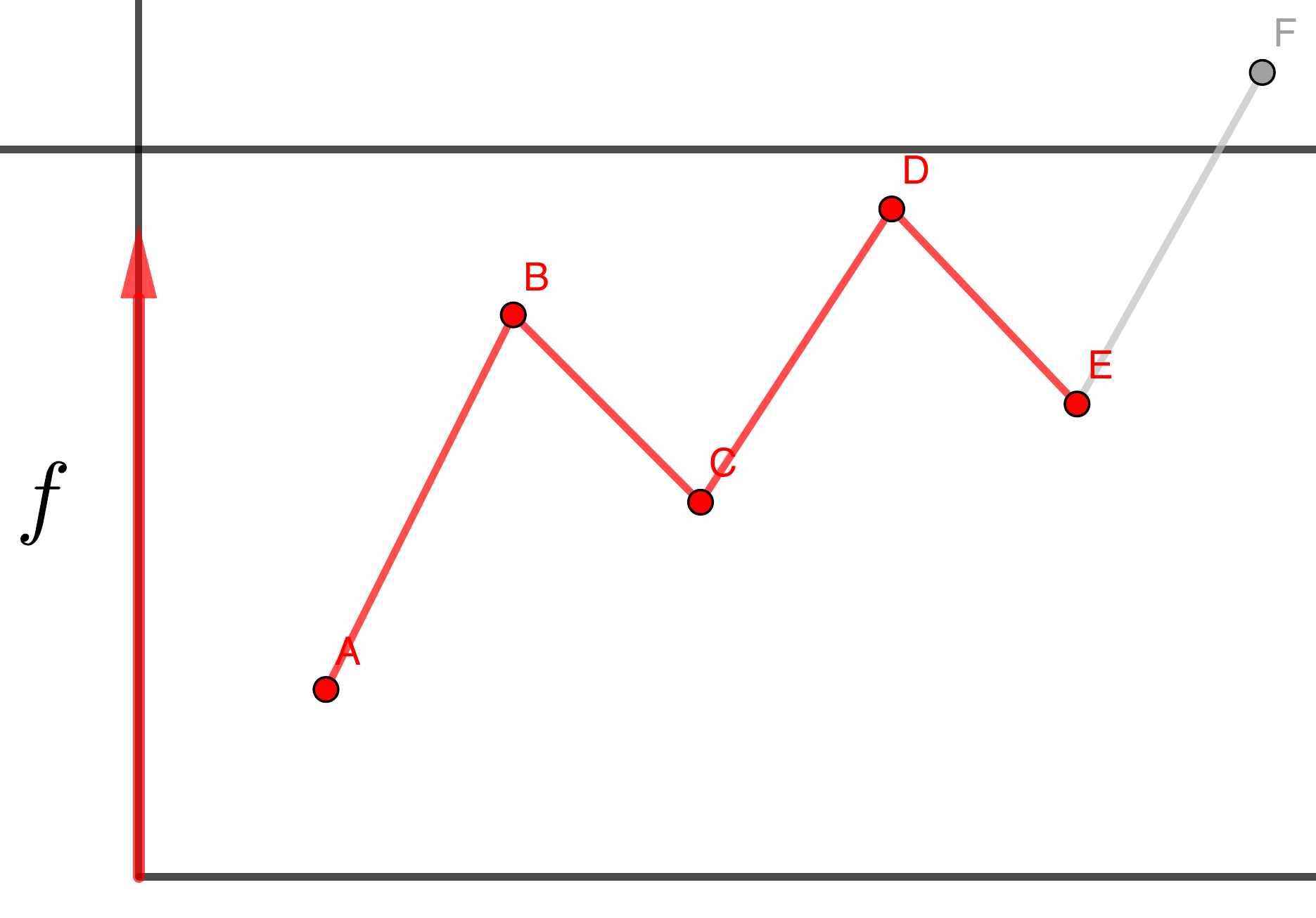} \hspace{10mm} & \includegraphics[width = 0.37 \linewidth]{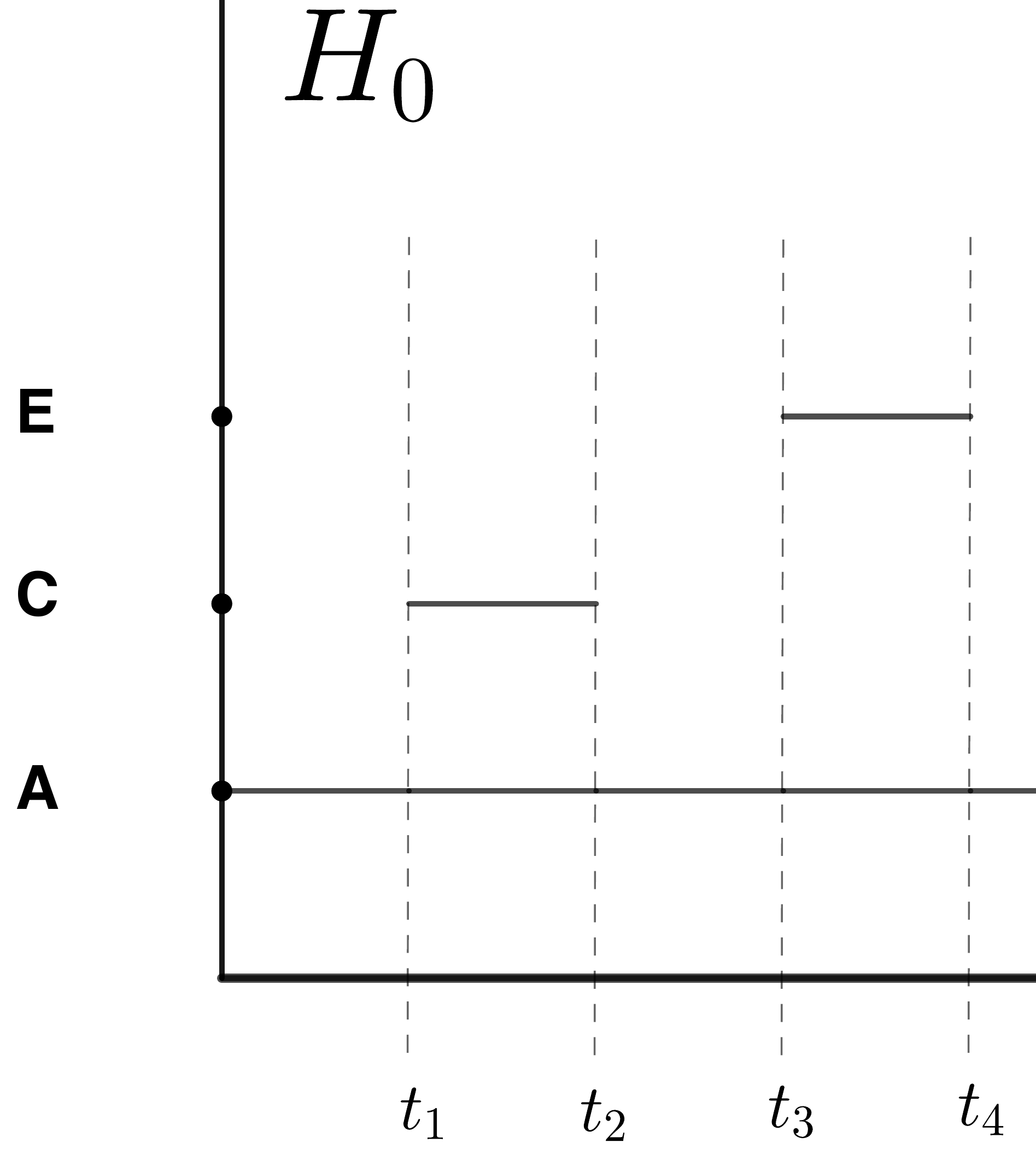} \hspace{5mm}\\
        \hline
    \end{tabular}
    \vspace{3mm}
    \caption{\textbf{Left:} Lower-star filtration. \textbf{Right:} The associated persistence barcode of 
the corresponding lower-star filtration pictured on the left.}
    \label{fig:lowexam}
\end{table}
\end{center}

\vspace{-10mm}

Once the lower-star filtration is obtained, persistent homology can be computed as follows. The inclusion $K_{i}\subset K_{j}$  induces a homomorphism $f^{i,j}:H(K_{i})\rightarrow H(K_{j})$  on homology. 
 Its  image is the persistent homology, letting $\beta_p^{ij}$ be the number of $n$-dimensional `holes' that are born at $K_i$ and die entering $K_j$. During the computation of persistent homology along the
filtration, an elder rule is applied. 
For example, when there are two connected components that get joined at some $K_j$, the older one (the one that was born earlier) remains, and the younger one dies. A persistence barcode is a representation of births and deaths of homology classes
%components 
along time using bars. An example is shown in Fig. \ref{fig:lowexam}.

Finally, once persistence barcodes are obtained, persistent entropy can be computed.
%based on Shannon entropy. %on a barcode.

\begin{definition}[Persistent entropy \cite{RUCCO2017130}]
\label{th:persentr}
Given a filtered simplicial complex $\{K(t) \ : \ t \in F\}$, and the corresponding persistence barcode $B = \{a_i = [x_i,y_i) \ : \ i \in I\}$, the persistent entropy $E$ of the filtered simplicial complex is calculated as follows:
$$ E = -\sum_{i\in I} p_i log(p_i)$$
where $p_i = \frac{l_i}{L}, l_i = y_i-x_i$, and $L= \sum_{i\in I}l_i$. In the case of an interval with no death time, $[x_i , \infty)$, the corresponding barcode $[x_i , m)$ will be considered, where $m = \max{\{F\}} + 1$.
\end{definition}

%%%%%%%%%%%%%%%%%%%%%%%%%%%%%%%555
%ROCIO: Aquí hace falta un ejemplo a partir del calculo de los barcodes del ejemplo anterio
%%%%%%%%%%%%%%%%%%%%%%%%%%%%%%%%%%%%%%%%%

The robustness of persistent homology  to noise is guaranteed
thanks to the following result,
letting a stable comparison between signals.
%as we will see later. 

\begin{theorem}
[\cite{RUCCO2017130}]
\label{thcontinuity}
Given two functions, $f : V \rightarrow R$ and $g : V \rightarrow  R$, defined on a set of vertices $V$ of $\mathbb{R}^n$, then for every $\varepsilon > 0$, there exists $\delta > 0$ such that
$$||f - g||_\infty \le \delta \Rightarrow | E(f) - E(g) | \le \varepsilon.$$
\end{theorem}

\medskip \noindent {\bf Machine Learning Background.} Machine learning techniques are nowadays widely applied to solve classification problems.

A classification technique will use a `training' dataset 
$$D=\{ \ (\vec{v}_i,c_i) \ | \ \vec{v}_i \in \mathbb{R}^n, \ c_i\in \{0,\dots,k\}, \ i\in \{1,\dots, m\} \ \}$$ 
where $\{0,\dots,k\}$ are the different possible classes. From this dataset, the classification algorithm will produce a classification model. This model can lately be applied to new inputs in order to predict the corresponding classes. There exist several classification techniques in machine learning. In our case, we focus our attention on support vector machine (see
\cite{boser}, \cite{cortes1995},
\cite{cristianini2000introduction} 
and \cite[Chapter 5]{geron2017hands-on}.

%For a deeper explanation than the one we are providing here about support vector machines, \cite{cristianini2000introduction} and \cite{géron2017hands-on} chapter 5 can be consulted. 

A support vector machine is a supervised learning technique that construct a hyperplane, driven by a linear function $b+\sum_{i=1}^m \alpha_i \vec{v}_i^T\vec{v}_i$,
or a set of them that can be used to classify data. When this data is not linearly separable, a kernel trick is applied: the space is mapped to higher dimensions using a kernel function, 
$k(\vec{v},\vec{v}')=\phi(\vec{v})^T\cdot \phi(\vec{v}')$.
Therefore, a support vector machine just creates hyperplanes that work as decision boundaries for classification after applying a deformation of the dataset in order to get a linearly separable representation. Then, formally, a support vector machine within a kernel makes predictions using the following function: $$f(\vec{v}) =  b+\sum_{i=1}^m \alpha_i k(\vec{v},\vec{v}_i)$$
where $\alpha$ is a vector of coefficients, $k$ the kernel and $b$ is a bias term. Finally, the coefficients are chosen as a result of an optimization problem of the separation margin between classes. Different kernel-based functions can be used, for example:

\begin{table}[]
\centering
\begin{tabular}{|c|c|}
\hline
\multicolumn{2}{|c|}{\cellcolor[HTML]{FFFFFF}{\color[HTML]{000000} \textbf{Kernels}}}                            \\ \hline
Linear                   & $k(\vec{v},\vec{v}') = \vec{v}^T\cdot \vec{v}'$                             \\ \hline
Polynomial of degree $d$ & $k(\vec{v},\vec{v}')=(\vec{v}^T\cdot \vec{v}'+c)^d $                        \\ \hline
Gaussian                 &  $k(\vec{u},\vec{v})= \mathcal{N}(\vec{u}-\vec{v};0,\sigma^2\vec{I})$ \\ \hline
\end{tabular}
\end{table}
\noindent where $\mathcal{N}(\vec{v};\vec{\mu},\Sigma)$ is the standard normal density.

\medskip \noindent {\bf Performance Metrics.} Basically, we are dealing with a classification problem. Therefore, our main metric will be the \textbf{accuracy}, considered as the percentage of well classified data in a dataset:

$$\mbox{Accuracy} = \frac{m}{n} $$
where $m$ is the number of well-classified data and $n$ is the size of the full dataset used in the test.
%%%%%%%%%%%%%%%%%%%%%%%%%%%%%%%%%%%%%%%%%%%
%ROCIO: precision and recall no se define ni se da una referencia. Ademas si se pone la frase debería ser en future work y no aquí.
%For further developments over this initial study we will introduce additional metrics such as \textbf{precision} and \textbf{recall}.
%%%%%%%%%%%%%%%%%%%%%%%%%%%%%%%%%%%%%%%%%%%%%%%%%%%%%%%%%

\medskip \noindent {\bf Statistical Tool.} The correlation coefficient of two random variables is a measure of their linear dependence. One correlation coefficient largely known and applied is the Pearson's correlation coefficient \cite{pearson1895note}:
$$\mbox{Pearson's correlation coefficient }\rho(A,B)=\frac{cov(A,B)}{\sigma_A \sigma_B} $$
where $cov(A,B)$ is the covariance and $\sigma$ the standard deviation. 

%%%%%%%%%%%%%%%%%%%%%%%%%%5
%ROCIO: no creo que esto necesite un entorno definition al igula que Accuracy tampoco lo necesitaba
%\begin{definition}[Correlation coefficient %\cite{pearson1895note}]
%The Pearson's correlation is given by the following %expression
%$$\rho(A,B)=\frac{cov(A,B)}{\sigma_A \sigma_B} $$
%where $cov(A,B)$ is the covariance and $\sigma$ the standard %deviation. 
%\end{definition}
%%%%%%%%%%%%%%%%%%%%%%%%%%%%%%%%%%%%%%%%%%%%

\section{Methodology}\label{sec:methods}

As was previously anticipated, the shape of the wave of a speech signal can be meaningful to emotional speech recognition. Roughly speaking, 
we
will compute persistent entropy to the lower-star filtration of the raw signal and then, we classify the signals by comparing these numbers using a support vector machine. 

\begin{figure}
\centering
\includegraphics[width = \linewidth]{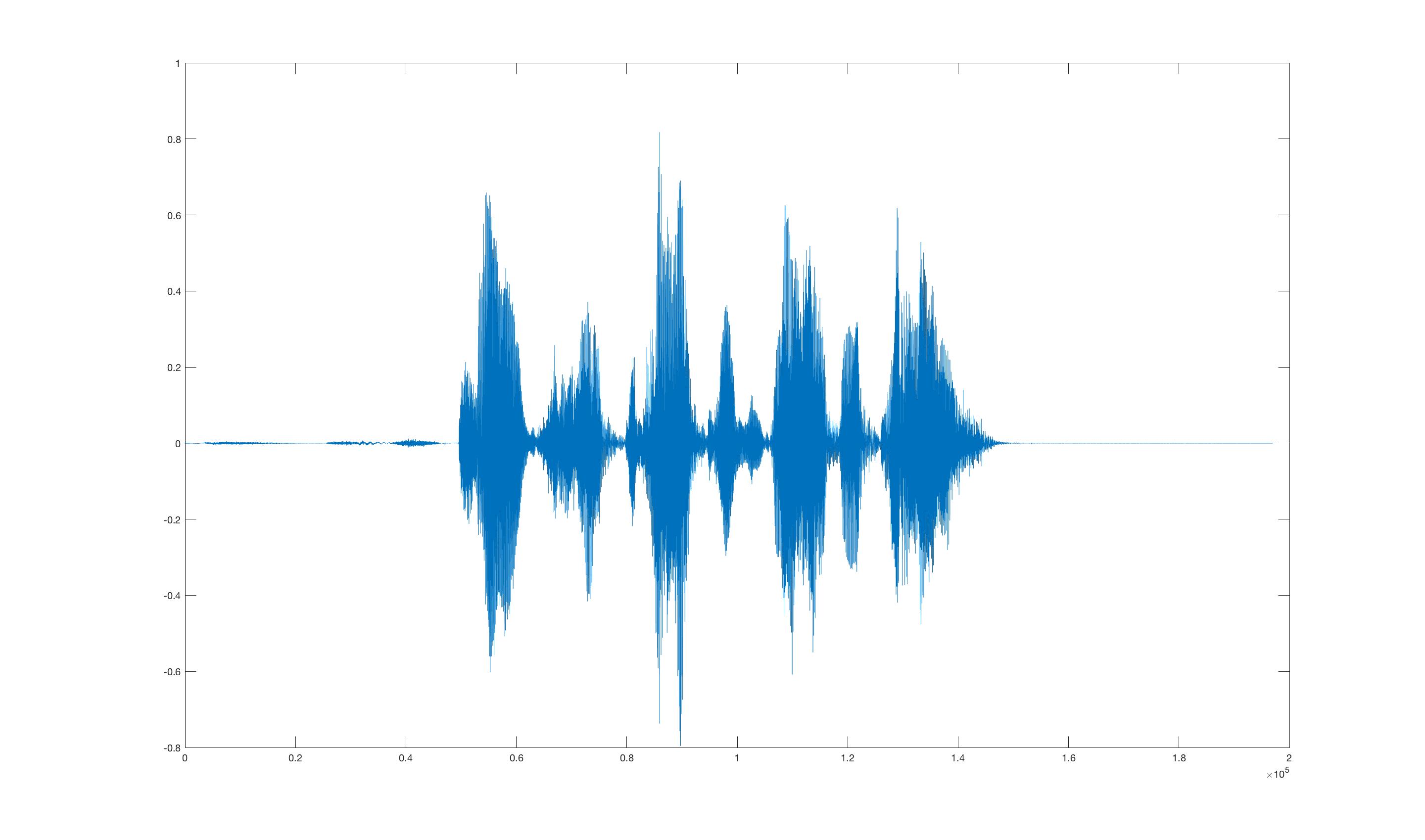}
\caption{Raw signal intensity graph of an
%. Concretely, 
angry emotion interpreted by the actor number 1 
of the RAVDESS dataset.}
\label{fig:senal33}
\end{figure}

Let us now explain in details the methodology applied in this paper:
%deeper the methodology applied in this paper by steps:

\medskip \noindent \textbf{Step 1.
Subsampling of the signal.} The size of each signal
%Given a signal, the size 
is reduced in order to face the complexity of the persistent homology algorithm. Besides, every signal of the dataset needs to be subsampled into the same size in order to fulfill the assumptions of Theorem \ref{thcontinuity}.  For example, %given the 
we subsampled the signal pictured in Fig. \ref{fig:senal33}
%following signal (see Fig: \ref{fig:senal33}),  it 
from 196997 points to 10000. The subsampling process was done uniformly on the signal, maintaining its shape and main distribution of the spikes. Furthermore, the experiments of Section \ref{sec:experiments} were also done using the dataset without subsampling reaching similar results. Then, we could assert that
%it can be considered that 
this type of subsampling does not loose relevant information for this approach.

\medskip \noindent  \textbf{Step 2. Introduction of imperceptible noise.} 
Signals are slightly perturbed  to fulfill the requirement of lower-start filtrations (see Definition \ref{def:starfiltr}): two points in the signal can not have the same height.
%\cR{Since  the lower-start} filtration requires that every point has a different height (see Definition \ref{def:starfiltr}), imperceptible noise is introduced to \cR{signals to fullfill the requirement}
%those points that require it. 

\begin{figure}[h!]
\centering
\includegraphics[width = \linewidth]{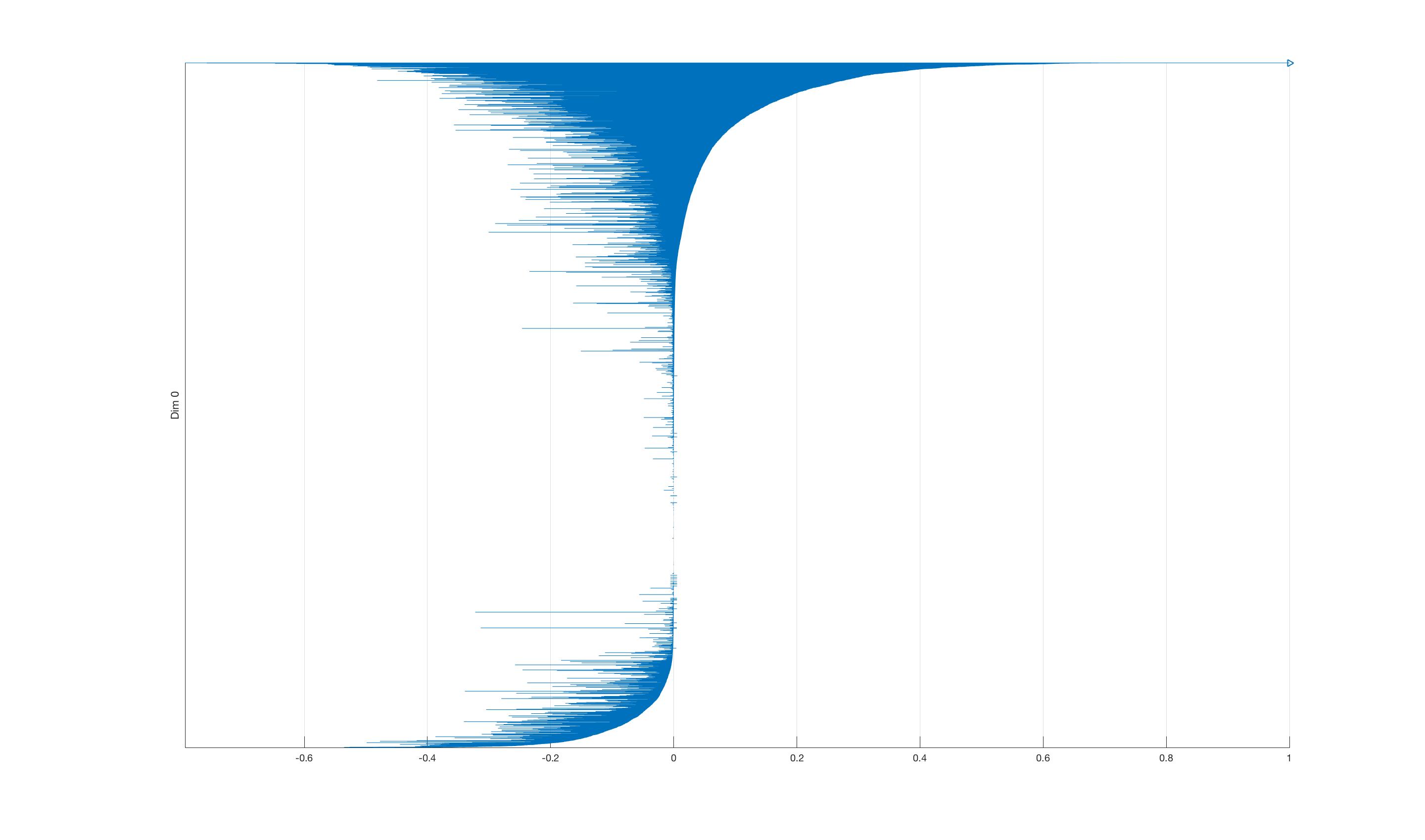}
\caption{Barcode of the signal shown in Fig. \ref{fig:senal33}. The horizontal axis represents %the 
time. Every horizontal (blue) line represents the life of a 0-dimensional homology class.}
\label{barcode33}
\end{figure}

\medskip \noindent  \textbf{Step 3. 
Persistence barcode computation.}
%Application of the lower-star filtration to points and extraction of the associated persistent barcode.} 
The lower-star filtration technique is applied to the 
signals
%results 
generated in  Step 2, obtaining the associated persistence barcode. For example, the barcode associated to the signal of Fig. \ref{fig:senal33} can be seen in Fig. \ref{barcode33}.

\medskip \noindent  \textbf{Step 4. Persistent entropy computation.} Persistent entropy is computed applying the formula given in Definition \ref{th:persentr} to the persistence barcodes obtained in Step 3. 

\medskip \noindent  \textbf{Step 5. Support vector machine classification.} This step consists of the application of several support vector machines with different kernels in order to infer results and develop a classification predictor to emotions. The different possible kernels, previously introduced in the paper, are tested and the one with better accuracy is chosen.

\section{Experiments}\label{sec:experiments}
The work-flow presented in the previous section was applied to the RAVDESS dataset \cite{ravdess}. This dataset is composed by $24$ actors interpreting $60$ audios each on different emotions and different intensity. Concretely,
there are $4$ audios for the neutral emotion and $8$ audios for each of the seven remaining emotions. Consequently, there are $1440$ different audios. 

\begin{figure}
\centering
\includegraphics[scale = 0.8,angle = 90]{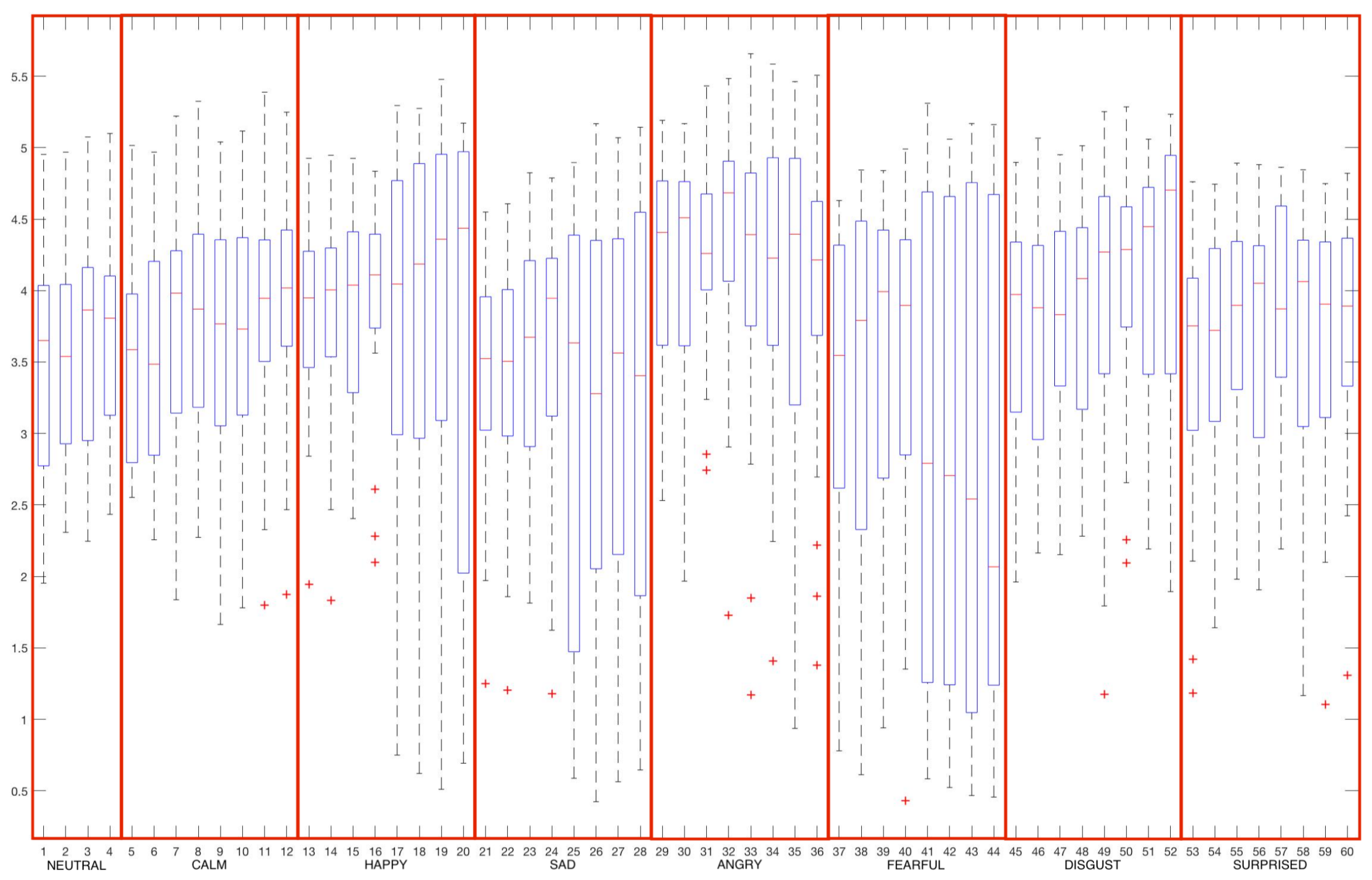}
\caption{(90\degree \ rotated figure) Horizontal axis represents the different $60$ audios. Vertical axis represents  persistent entropy.
The big (red) rectangle clusters encloses  persistent entropies of the audios per emotion (the respective emotion is indicated in the horizontal axis.
The small (blue) rectangles are quartiles for the persistent entropy values. The vertical (blue) dashed lines  mean the range of values of persistent entropy values. The (red) points are outliers. The horizontal (red) small lines are the mean persistent entropy value for the corresponding audio.}
\label{boxplotfig}
\end{figure}

In Fig. \ref{boxplotfig}, a box-plot of the persistent entropy of the $1440$ audios grouped by the different emotions
%emotions 
can be seen. 
%From it, 
We
can infer that persistent entropy values vary depending on
both
the emotion and the person. It seems that there exists characteristic personal values and
%. Furthermore, 
the range of every emotion can be really wide. For example, the persistent entropy values of the 
audio number $20$ in Fig. \ref{boxplotfig}, that is an example of happiness, varies from $5.1713$ to $0.6923$ depending on the person. Besides, the existing overlapping between the boxes tells us
%let us know 
that %all 
emotions can not be distinguished from the rest by just the persistent entropy values of every script as a feature. This failure approximation is illustrated and explained in Experiment 1. However, 
some emotions
%some of the emotion 
can be differentiated by pairs even with this `naive' approximation. 

\begin{figure}
\centering
\includegraphics[scale = 0.25]{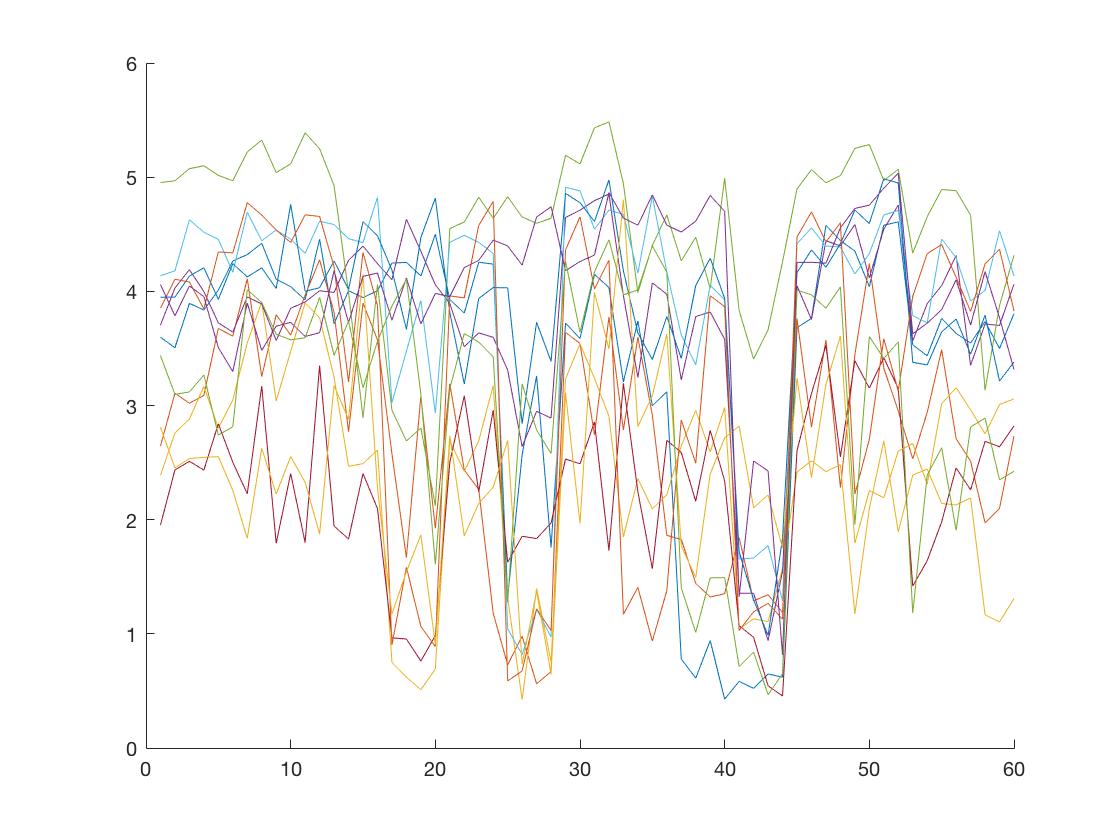}
\caption{Horizontal axis represents  the different audios of actresses. Vertical axis represents persistent entropy value. 
The different persistent entropy values for the 60 audios of the same actress are connected by an straight line.
%Linear unions of the different persistent entropy values for the 60 audios per female actor. Each linear union belong to a different actress. 
%
We can see that  shapes are correlated (see Table \ref{tab:correlation}), showing that they tend to have the same peaks and downs.}
\label{female}
\end{figure}

\begin{figure}
\centering
\includegraphics[scale = 0.25]{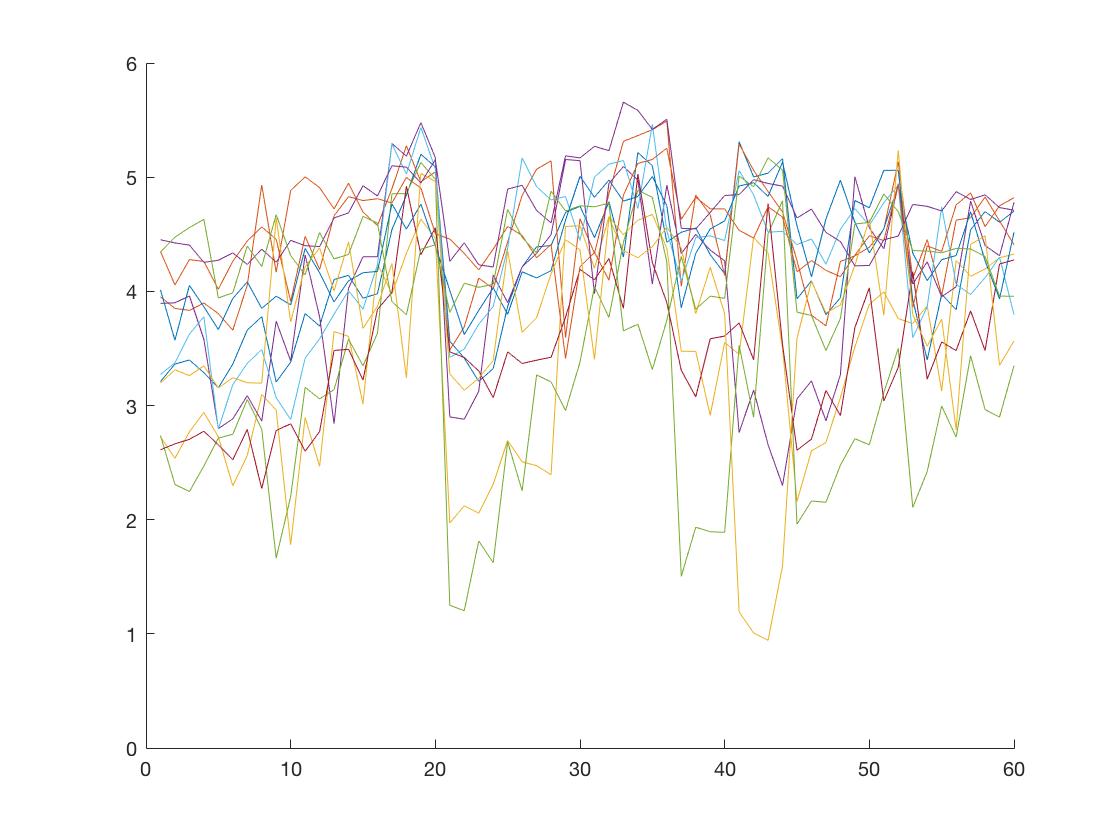}
\caption{
Horizontal axis represents  the different audios of male actors. Vertical axis represents persistent entropy value. 
The different persistent entropy values for the 60 audios of the same  actor are connected by an straight line.
%Horizontal axis mean the different audio. Vertical axis is the persistent entropy value. Linear union of the different persistent entropy values for the 60 audios per male actor. Each linear union belong to a different actor.
We can see that  shapes are correlated (see Table \ref{tab:correlation}), showing that they tend to have the same peaks and downs.}
\label{male}
\end{figure}

One thing that appealed our attention is the visual correlation that persistent entropy values tend to have per sexes as shown in Fig. \ref{female} and Fig. \ref{male}. Even if the range is lower or higher depending on the person, in general, the peaks appear on the same places. 
To illustrate it, let us consider the correlation matrix between 
persistent homology values of
the $60$ audios grouped in the ones belonging to females and the ones belonging to males.
%per person and let us compare \cR{persistent homology values between females and males.}
%
We obtain that persistent entropy values
%vectors 
are moderately correlated between same sex audios and badly correlated between different sexes 
(see Table \ref{tab:correlation}). 
%%%%%%%%%%%%%%%%%%%%%%%%%%%%%%%%%%%%5
%ROCIO: no entiendo esta frase
%It bodes well for the existence of a significant case of study 
%%%%%%%%%%%%%%%%%%%%%%%%%%%%%%%%%%%%5
 We think that it could be interesting the use of more 
 sophisticated
 %expressive 
 measures of similarity apart from correlation. Furthermore, correlation results give
 %it  gives 
 us clues to the need of developing emotion classification within the dataset separated by sexes to reach better classification accuracy. Besides, we consider that persistent entropy values could even be a nice approach to people identification and not just to emotion recognition. However, this approach is far from the scope of this paper and its preliminary nature.

\begin{table}
\centering
\begin{tabular}{|l|c|c|}
\hline 
       & Male actor & Female actor \\\hline
Male actor   & 0.43     &   0.23     \\\hline
Female actor & 0.23     &   0.49   \\ \hline 
\end{tabular}
\vspace{2mm}
\caption{Mean values for the correlation coefficients of the entropy values grouped
%vectors 
by sexes.}
\label{tab:correlation}
\end{table}
In all the following experiments we use as the classification technique a support vector machine with fold cross validation and the kernel that provides the better accuracy from the ones explained previously. The training dataset will be the $1440$ persistent entropy values grouped by different ways trying to get the features needed to reach our goal. In the first experiment we try the brute force approach using every script as a point of the training dataset. Then, in the second experiment, every point correspond to an emotion within its $24$ persistent entropy values by the $24$ different actors. Finally, in the last experiment, the dataset is grouped by actors and emotions. 

\noindent {\bf Experiment 1:} Each persistent entropy value will be a point of the training dataset. In this case, $20.3\%$ of accuracy is reached within a linear kernel. Some conclusions can be pointed out from this failed approach: The emotion recognition problem is a multidimensional one, in the sense that a 1-dimensional embedding is not enough to 
an acceptable
%a nice 
classification result. Furthermore, this was anticipated by the overlapping of the different boxes at the box-plot of persistent entropy values showed in Fig. \ref{boxplotfig}. Besides, the non correlation between 
persistent entropy values per
sexes is a matter %that is 
not taken into account in this experiment. 

%As we can see in Fig. \ref{feelingrange}, no differences can be outlined using that approximation. 
%\begin{figure}
%\centering
%\includegraphics[width = %\linewidth]{Rangos_emociones}
%\caption{The range of values for emotions does not let us to differentiate them if we are not considering a sample within different actors. Neutral was not included because of having less audios in the dataset. Horizontal axis is the $7$ different emotions and the vertical axis is the entropy values.}
%\label{feelingrange}
%\end{figure}

\noindent {\bf Experiment 2:} Each point of the  dataset is a vector of $24$ features which correspond to the persistent entropy value of the same emotion interpreted
by the $24$ different actors. The dataset was separated in $40$ points for training dataset and $20$ points for test dataset and a gaussian kernel was used. Then, $92.5\%$ of accuracy was reached on the training dataset and $90\%$ on the test dataset. Furthermore, $96.66\%$ accuracy was obtained on the full dataset. In our opinion, this experiment presents two main drawbacks. The first one is its difficult applicability  as it needs %these 
$24$ features of every emotion. However, withing long audio recordings, it could be cut into pieces and obtain enough features to classify. The other drawback is the small dataset we have for this experiment because of the way it has been grouped. 
%However, even if this result look great, this does not seem really applicable as we would need for new predictions, the same audio registered by different people feeling the same.  Even though it has given us the certainty that persistent entropy values are not able to generalize for every person and different casuistic need to be considered per person or in subgroups of population that share same characteristics. 

\medskip \noindent {\bf Experiment 3:} In this experiment, each point of the dataset consists of a vector of $8$ features, corresponding each feature to the persistent entropy value of the same emotion interpreted by the same actor. By this, the following accuracy Table \ref{table:accuracy} for classification by pair of emotions was obtained using a second degree polynomial kernel. 
Considering other results in the literature %from other authors 
like  \cite{soundresearch} where $71\%$ of accuracy was reached using Artificial Neural Networks, 
our results are really promising.
%%%%%%%%%%%%%%%%%%%%%%%%%5%
%ROCIO: Aquí no aparece ningún dato de accuracy, por eso lo he quitado. PEros i lo añades, se puede volver a poner.
%or like \cite{YANG20101415} where pair classification was also done using other features of audios like harmonics.
%%%%%%%%%%%%%%%%%%%%%%%%%%%%%%%%%%%%%%%%%%%%%%%%%
However, we are still far from the $83\%$ of accuracy
reached in \cite{zhang} using a multi-task hierarchical model. 
But we can say that, with just a first approximation, we could reach similar accuracy than those that already exist in the literature.
%obtained by other authors. 
Furthermore, as we are considering here just intensity and one type of filtration, only some features that characterize emotions are taken into account. Then, it gives us a nice starting point in order to improve the model
by  using different features of the signal and different filtrations.
%at the \cR{persistent} homology algorithm. 
%%%%%%%%%%%%%%%%%%%%%%%%%%%%%%%5
%ROCIO: Esta frase no se entiende:
%Besides, this type of classification is more applicable because of when emotion classification is done, it is intended during a conversation and, then, it is easy to obtain several audios or one audio long enough (to be cut) in order to do predictions. 
%%%%%%%%%%%%%%%%%%%%%%%%%%%%%%%%%%%%%%

\begin{table}[]
\centering
\begin{tabular}{llllllll}
\cline{1-1}
\multicolumn{1}{|l|}{Feelings} & Calm                     & Happy                    & Sad                      & Angry                    & Fearful                  & Disgust                  & Surprised                \\ \cline{1-1}
Calm                           & \cellcolor[HTML]{C0C0C0} & 77.1\%                   & 68.8\%                   & 81.2\%                   & 79.2\%                   & 72.9\%                   & 60.4\%                   \\
Happy                          & \cellcolor[HTML]{C0C0C0} & \cellcolor[HTML]{C0C0C0} & 62.5\%                   & 64.6\%                   & 60.4\%                   & 58.3\%                   & 64.6\%                   \\
Sad                            & \cellcolor[HTML]{C0C0C0} & \cellcolor[HTML]{C0C0C0} & \cellcolor[HTML]{C0C0C0} & 75\%                     & 62.5\%                   & 70.8\%                   & 60.4\%                   \\
Angry                          & \cellcolor[HTML]{C0C0C0} & \cellcolor[HTML]{C0C0C0} & \cellcolor[HTML]{C0C0C0} & \cellcolor[HTML]{C0C0C0} & 68.8\%                   & 77.1\%                   & 70.8\%                   \\
Fearful                        & \cellcolor[HTML]{C0C0C0} & \cellcolor[HTML]{C0C0C0} & \cellcolor[HTML]{C0C0C0} & \cellcolor[HTML]{C0C0C0} & \cellcolor[HTML]{C0C0C0} & 72.9\%                   & 72.9\%                   \\
Disgust                        & \cellcolor[HTML]{C0C0C0} & \cellcolor[HTML]{C0C0C0} & \cellcolor[HTML]{C0C0C0} & \cellcolor[HTML]{C0C0C0} & \cellcolor[HTML]{C0C0C0} & \cellcolor[HTML]{C0C0C0} & 75\%                     \\
Surprised                      & \cellcolor[HTML]{C0C0C0} & \cellcolor[HTML]{C0C0C0} & \cellcolor[HTML]{C0C0C0} & \cellcolor[HTML]{C0C0C0} & \cellcolor[HTML]{C0C0C0} & \cellcolor[HTML]{C0C0C0} & \cellcolor[HTML]{C0C0C0}
\end{tabular}
\vspace{2mm}
\caption{Prediction accuracy from pair of emotions using different support vector machine within different kernels. }
\label{table:accuracy}
\end{table}

\section{Conclusions and future work}
\label{sec:conclusions}
A persistent entropy application has been developed in order to extract information from raw audio signals and solve a classification problem using support vector machine. Furthermore, a descriptive analysis of the computed
persistent entropy values has been done, bringing up the characteristic values that exist by person and the existence of moderate correlation between 
persistent entropy values of emotions of
people of the same sex. Additionally, we have provided insights showing that separating the dataset by sexes would get better accuracy for the classification task.  Finally, three different experiments have been proposed: two of them can be considered successful. This makes evidence that topological data analysis tools are a nice approach to this task, being interesting the development of more sophisticated algorithms. 

In this first approximation just $\beta_0$ has been used. However, there exists different processing techniques to signals that can obtain images from them and that would allow us to consider
higher dimensional topology features that can be meaningful for the emotion recognition task. We could combine them to reach a better prediction skill.

Another interesting approach is training the machine learning classification tool with the audios interpreted by just one actor, obtaining a personal trained emotion predictor. However, RAVDESS
%we consider that the 
dataset is not big enough to obtain interesting conclusions within this approach. Therefore, this would be a nice future 
work,
%continuation of the research, 
in these days that it is quite easy to obtain lot of data from users.

\begin{figure}
\begin{center}
\includegraphics[width = 0.6 \linewidth]{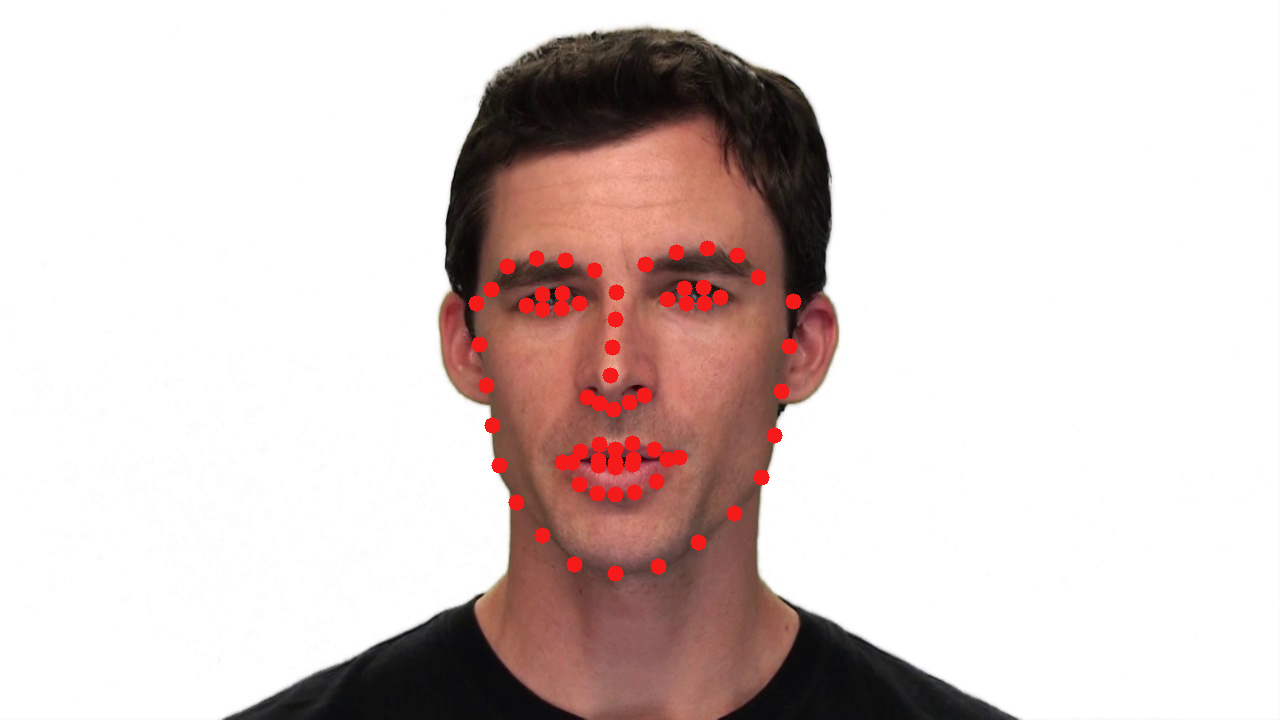}
\caption{Landmarks points of one frame of a video of the RAVDESS dataset.}
\label{fig:landmarks}
\end{center}
\end{figure}

Furthermore, as the associated videos of the audios are available in the RAVDESS dataset, we would like to use the landmarks (see Fig. \ref{fig:landmarks}) as input to topological data analysis tools (like a Vietoris-Rips filtration) and combine this information within the one provided by the audios used in this paper. Similarly, one of the most relevant conclusions that KRISTINA project reached was that the combination of visual and audio features can develop better predictions than using them separately. 

%%%%%%%%%%%%%%%%%%%%%%%% 
\medskip
\bibliographystyle{abbrv}
\bibliography{biblio.bst}

\end{document}